\documentclass[aps,preprint,showpacs]{revtex4}
\usepackage{graphicx,epsfig,pstricks,rotating}

\begin{document}

\title{Detrended Cross-Correlation Analysis: A New
Method for Analyzing Two Non-stationary Time Series
}

  \author{Boris~Podobnik}

  \affiliation{Department of Physics, Faculty of Civil Engineering, University of Rijeka,
    Rijeka, Croatia}
  \affiliation{Zagreb School of Economics and Management,
 Zagreb, Croatia}

  \author{H.~Eugene~Stanley}
  \affiliation{Center for Polymer Studies and Department of
    Physics, Boston University, Boston, MA 02215}

\begin{abstract} 
  Here we propose a method, based on detrended  covariance  
   which  we call  detrended cross-correlation
   analysis (DXA),  
  to investigate  power-law cross-correlations between different
  simultaneously-recorded time series in the 
  presence of non-stationarity. 
  We illustrate  the method  by selected examples  from physics, physiology,
  and finance. 
\end{abstract}
  
  \maketitle

\date{today}
There are a number of situations where different signals exhibit
cross-correlation.
 In seismology, the degree of cross-correlation among noise
 signals taken at different antennas
 of detector arrays is an alert signalling earthquakes 
 and volcanic eruptions \cite{Seismology}.
In finance,  risk  is estimated on the  basis of cross-correlation
matrices for different
 assets and investment portfolio  \cite{Finance}. In nanodevices for quantum information
processing, electronic entanglement motivates the computation of
current noise cross-correlations, to see whether the sign of this
signal would be reversed compared to the standard devices
\cite{Samuelsson}.

 Consider two time series  $\{y_i\}$ and  $\{y'_i\}$,
 where $i=1,2,..,N$.   Each series can be represented as a 
 random walk  of $k$  steps, and we can define 
  $R_k \equiv y_1 + y_2 + ... + y_k$    and
   $R'_k \equiv y'_1 + y'_2 + ... + y'_k$, where $k \le N$. 
  Series 
 $\{y_i\}$ has a  mean $\mu \equiv \overline{y_i} \equiv (1/N)
  \sum_{i=1}^N y_i$
  and a variance  $ \sigma^2 \equiv \overline{(y_i - \mu)^2}$, 
  while series $\{y'_i\}$ has a mean  
  $\mu' \equiv \overline{y'_i}$
  and a variance  $ \sigma'^2 \equiv \overline{(y'_i - \mu')^2}$. 
  We assume that the auto-correlation functions 
  ${\it A(n)} \equiv {\overline{(y_k - \mu) (y_{k+n} - \mu)}}
 / \sigma^2$ and  ${\it A'(n)} \equiv {\overline{(y'_k - \mu') 
 (y'_{k+n} - \mu')}} 
 / \sigma'^2 $ scale as power laws 
  $A(n) \sim n^{- \gamma}$  and
  $A'(n) \sim n^{- \gamma'}$,   with $ 0 < \gamma,  \gamma' < 1$. We further 
  assume that 
for the cross-correlation function 
$ {\it X(n)} \equiv {\overline{(y_k - \mu) (y'_{k+n} - \mu')}}
 /(\sigma \sigma') $ between the time series 
 $\{y_i\}$ and  $\{y'_i\}$  
\begin{equation}
X(n) \sim n^{- \gamma_{\times}}, 
\end{equation}
 with $ 0 <   \gamma_{\times} <1$.  However, 
this definition assumes {\it stationarity}  of both  time series, and 
one can question  its applicability  to real-world data typically 
characterized by a high degree of {\it non-stationarity}.

 Currently there is no method  to quantify  the cross-correlations
 exponent $\gamma_{\times}$ 
 between two correlated time series  in the presence  of non-stationarity
 \cite{korea}. 
  Here we propose such a method, and we illustrate  the method by selected 
  examples  from physics, physiology and  finance. 
 To this end, we  calculate the covariance:
   \begin{equation}
   \label{COVARIANCE}
  \overline{(R_n - \overline{R_n}) (R'_n - \overline{R'_n})}~ =
   n ~X(0) + 2 \sum_{k=1}^{n-1} [n X(k) -   k X(k)],
\end{equation}
 where 
 $X(0) =  \overline{(y_{k} - \mu) (y'_{k} -  \mu')} /\sigma \sigma'$. 
The sums  of Eq.~(\ref{COVARIANCE})  can be approximated
 by  integrals: $
 \sum_{k=1}^{n-1} X(k) \approx \sum_{k=1}^{n}  k^{-  \gamma_{\times}}
 \approx \int_1^n  dx ~ x^{-  \gamma_{\times}} \propto n^{1 -  \gamma_{\times}}$
  and $
 \sum_{k=1}^{n-1} k ~X(k) \approx \sum_{k=1}^{n}  k^{1-  \gamma_{\times}}
 \approx \int_1^n  dx~  x^{1-  \gamma_{\times}} \propto n^{2 -  \gamma_{\times}}$.
Asymptotically  Eq.~(\ref{COVARIANCE}) scales as 
 \begin{equation}
\overline{(R_n - \overline{R_n}) (R'_n - \overline{R'_n}) } ~
 \sim n^{2 \lambda},
\end{equation}
where the scaling exponents $\lambda$ and $\gamma_{\times}$  ---
respectively related to the covariance and the cross-correlation function ---
are not independent, since 
 $\lambda \equiv 1 - 0.5 \gamma_{\times}$ \cite{power}. 
 For $\{y_i\}= \{y'_i\}$, the covariance of  Eqs.~(2)-(3)  becomes 
 the variance that for $n>> 1$  scales as
$ n^{2 H} $, so $\lambda=H$, where $H$ is the Hurst
 exponent.

In order to quantify  long-range cross-correlations when
non-stationarities are present, we  propose a modification of the
above covariance analysis which we call {\it detrended
 cross-correlation analysis} (DXA).
We consider two long-range cross-correlated time series $\{y_i\}$ and
$\{y'_i\}$ of equal length $N$, 
 compute two integrated signals  $R_k \equiv \sum_{i=1}^{k} y_i   $  and  
$R'_k \equiv \sum_{i=1}^{k} y'_i $, 
where $k=1,..,N$.  We divide the entire time series into $N-n$  overlapping
boxes, each containing $n+1$ values. For both time series,
in each box that starts at $i$ 
and ends at $i+n$, 
we define the "local trend"  to be the ordinate of a linear least-squares 
fit. We define the "detrended walk"  as the difference between 
the original walk and the local trend.  
Next we calculate the covariance of the residuals in each box
 $f^2_{DXA}(n,i) \equiv 1/(n-1) \sum_{k=i}^{i+n}  
  (R_k - {\widetilde{R_{k,i}} })
	(R'_k - {\widetilde{R'_{k,i}} })$. 
Finally, we calculate the   detrended covariance   by summing over all
overlapping $N-n$ boxes of size $n$, 
\begin{equation}
\label{DXA}
   F_{DXA}^2(n) \equiv \sum_{i=1}^{N-n} f^2_{DXA} (n,i).  
\end{equation}
 When only one random walk
is analyzed ($R_k=R'_k$),  the {\it detrended covariance}  $F^2_{DXA}(n) $ 
 reduces to the {\it detrended variance}  $F^2_{DFA}(n)$ used in the 
DFA method \cite{CKP}.


 In order to test the utility of the proposed DXA method, 
 power-law auto-correlated time  series
$y_i$ and $y'_i$   are generated by
 using a stationary linear
``ARFIMA'' process \cite{Hosk81}: 
 $y_i \equiv \sum_{j=1}^{\infty} a_j(\rho) y_{i-j} +
 \eta_i$ \cite{cutoff}, where
  $ 0 < \rho < 0.5$  is a free
  parameter, $a_j(\rho)$ are weights defined by $a_j(\rho) \equiv 
  \Gamma(j-\rho) / [\Gamma(-\rho) \Gamma(1+ j)]$, $\Gamma(j)$ denotes the
  Gamma function,  and $\eta_i$  denotes an independent and identically
  distributed (i.i.d) Gaussian variable. 
  The parameter $\rho$ is 
  related to the Hurst exponent, $H = 0.5 + \rho$ \cite{Hosk81}.
We generate two time series:
 $\{y_i\}$ with $\rho = 0.1$ and
  $\{y'_i\}$ with $\rho' = 0.4$.
 Since both  $\{y_i\}$ and $\{y'_i\}$
 are generated with the same error 
 term $\eta_i$, $X(n) \ne 0$ \cite{artificial}. 
 In Fig.~1(a)  we show that each time series exhibits
the power-law auto-correlations expected for ARFIMA, and 
  that the root mean square (rms) of the detrended covariance  vs.  $n$
  also follows approximately  a power law, consistent with the
  fact that   $\{y_i\}$ and
  $\{y'_i\}$      are power-law cross-correlated.
For different pairs 
 of power-law auto-correlated 
 time series $\{y_i\}$ and $\{y'_i\}$, 
 characterized by Hurst exponents $H$ and $H'$, 
we find  the time series are also 
 power-law cross-correlated, where the  exponent $\lambda$ 
 of Eq.~(3) 
 is approximately  equal to the average of the Hurst exponents:  
   $\lambda \approx (H + H')/2$.

 The power-law cross-correlations  between    
   $\{y_i\}$ and  $\{y'_i\}$ may exist only if $A(n) \sim n^{- \gamma}$ 
 for both processes.  
We generate  two time series by using two un-coupled ARFIMA
 processes: $\{y_i\}$ with $\rho =0.1$ and $\{y'_i\}$ with $\rho'=0.4$.
In Fig.~1(b) we find that, even though both $\{y_i\}$ and $\{y'_i\}$ are
 power-law auto-correlated, 
the detrended  covariance vs. $n$ of Eq.~(4) 
   fluctuates around zero which indicates that no power-law
    cross-correlations are present. The same result we  show
    for uncoupled ARFIMA processes $\{y_i\}$ and $\{y'_i\}$ defined by
      $\rho=0.2$ and $\rho'=0.3$, respectively.
Generally,  if the detrended
   covariance vs. $n$
  oscillates around zero, there are no power-law cross-correlations with
	 an unique exponent, but 
   either no cross-correlations or only 
  short-range cross-correlations exist between  $\{y_i\}$ and  $\{y'_i\}$. 

  To further exemplify the potential utility  of the DXA method for
  analyzing real-world data, we study two time series, both of which can be  
  considered as two outputs of a complex  system:   
 the air humidity  and  the  
  air temperature \cite{www}. We analyze 
  absolute values of the
 successive differences  of air humidity  (denoted by  
$\{|y_i|\}$) and 
air temperature (denoted  by $\{|y'_i|\}$) [see Fig.~2(a)]. 
  Fig.~2(b)
  shows that each of two  time series $\{|y_i|\}$ and $\{|y'_i|\}$  
  exhibits  power-law  autocorrelations with  similar scaling exponents.  
     Fig.~2(b)
 also shows that cross-correlations  
 between $\{|y_i|\}$ and $\{|y'_i|\}$ exist and can be fit a power 
law $n^{\lambda}$  with exponent $\lambda=0.75$, practically equal to the
exponent calculated for the temperature differences.  
   We also analyze time series $\{y_i\}$ and $\{y'_i\}$, and we find   
 that the DFA and DXA analyses exhibit the correlated behavior,    
where $F_{DXA}^2(n)$ is negative for every $n$ \cite{remark}. 
  
  As a second example of real-world data, we analyze the Sleep 
  Heart Health Study (SHHS) 
 database which is designed to clarify  the relationship    between 
 sleep disordered breathing and cardiovascular disease \cite{SHHS}.
 For a single patient \cite{SHHS,data},   
  we analyze  correlations behavior of five variables: 
  two electroencephalography 
  (EEG) variables, where EEG is  the   neuro-physiological  measurement
  of the electrical activity of the brain recorded by 
electrodes   commonly placed  on the scalp; 
	 heart rate (HR) describing the frequency of the cardiac
	 cycle, derived from  the electrocardiogram 
	(ECG) which records the electrical activity of the heart over time; 
 and for both left and right 
	 eye the   electrooculograms, obtained by measuring 
	 the resting potential of the retina. 
	 
	We analyze the time series of two EEG variables 
	simultaneously recorded every second, and  find that  each  of them 
	is short-range auto-correlated. 	 
	The absolute values of two EEG variables 
we show in Fig.~3(a). 
 The  DFA curves in Fig.~3(b) show that each time series of the magnitudes 
 exhibits 
power-law  auto-correlated  behavior, indicating that a large 
increment  is
   more likely to be followed by a large increment. 
Fig.~3(b) also shows  that, besides auto-correlations,
	 the time series of magnitudes exhibit power-law cross-correlations indicating 
	 that a large increment in one variable is more likely to be followed 
	 by large  increment in the other variable.
	 We also find that power-law magnitude  cross-correlations 
	 exist between the two time series of magnitudes of two EOG variables. 
	 We also find 
	 non-vanishing cross-correlations between ECG time series and   
	 	C3/A2 time series. 
In cross-sectional 
 studies where many different physiological time series are recorded, 
 an analysis based 
 on the  DXA method should  add diagnostic power to existing
 clinical methods employed to discriminate healthy from pathological behavior.

 As a third example we analyze   the daily closing values of the 
 Dow Jones and the Nasdaq financial 
 indices together with their corresponding trading volumes  
 (the number of  shares  traded  each day). For 
 both price and trading volumes, 
 we analyze the time series of absolute values of
  the differences of logarithms for successive days.  
  In  Figs.~4(a) and 4(c) we show their integrated signals
    $ I(n) \equiv \sum_{i=1}^n (|y_i| - \overline{|y_i|})$. 
	  In  Figs.~4(b)  and 4(d)
  we see  
  that each of the four time series  is power-law 
 {\it auto-correlated}, and we also see   both for absolute values of 
  price  changes (``volatility'') and trading volume 
  that the   
    time series   for Nasdaq and Dow Jones indices  are power-law 
	 {\it cross-correlated}. 

We obtain similar results by analyzing cross-correlations 
 between Microsoft and IBM stock  prices and trading volumes, 
	consistent with the interesting possibility  that the above 
	 results may  hold not only  for indices,  but also for individual
	  companies. 
	This result  is especially interesting
 during  volatile periods.  
 Long-range cross-correlations between two stocks imply that 
 each stock  separately  has  long memory   of its own 
 previous values, and also long memory 
  of previous values of  the other stock.

	Note that it is always possible that cross-correlations 
	between two time series are only apparent and exist only 
	due to the presence of long-range auto-correlations in 
	separate time series. 
	To test that cross-correlations in magnitudes between two 
	different 
	time series  are genuine,  we generate an artificial  time series $\{x_i\}$
	that is strongly auto-correlated in the magnitudes $\{|x_i|\}$.  
	We find  that there are no cross-correlations with an unique
	power-law  exponent between 
	$\{|x_i|\}$  and any of the empirical  time series analyzed in the paper.

 We thank the Ministry of Science of Croatia, and 
 the NSF for financial support.

  \vspace*{0.cm}

\vspace*{-1.9cm}


 \begin{figure}
    \centerline{
      \epsfysize=0.5\columnwidth{{\epsfbox{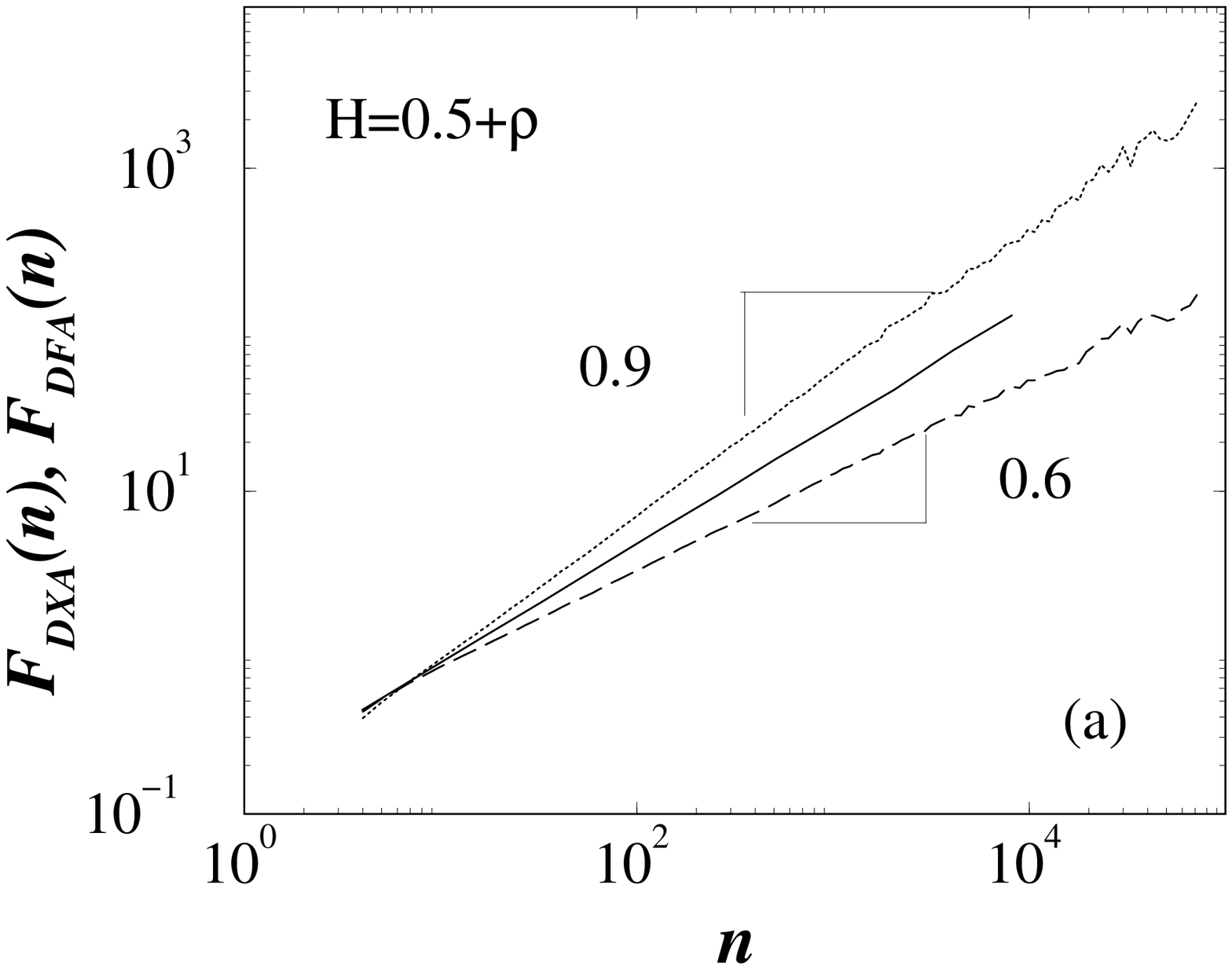}}}    }
  \vspace*{-.5cm}
  \end{figure}
\begin{figure}
    \narrowtext
    \centerline{
      \epsfysize=0.5\columnwidth{{\epsfbox{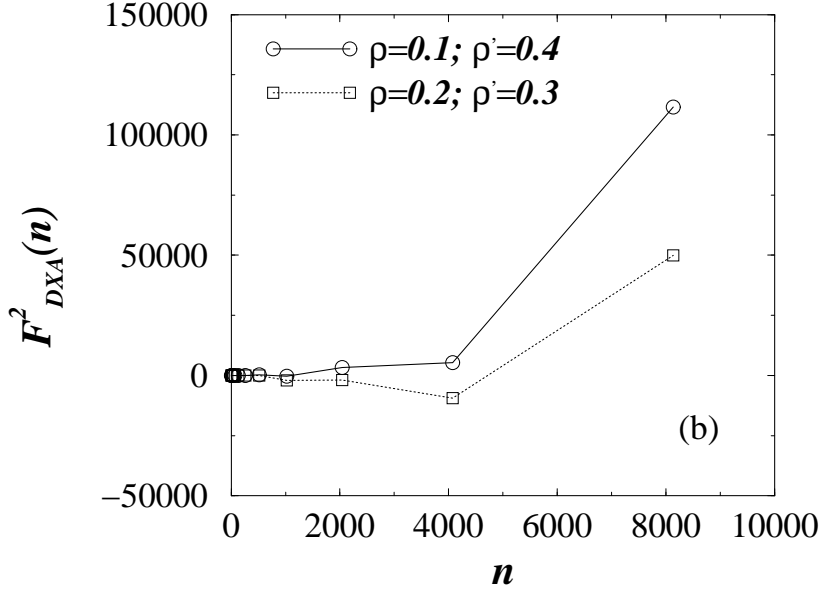}}}}
\caption{(a) Rms of detrended variance $F_{DFA}(n)$ and detrended 
covariance, $F_{DXA}(n)$, where $n$ is a scale.  
For two time series  generated  by two
 ARFIMA processes: $\{y_i\}$ with  $\rho =0.1$  and $\{y'_i\}$ 
  with $\rho'=0.4$
 we  show the DFA curves $F_{DFA}(n)$ for both  $\{y_i\}$ and $\{y'_i\}$,
which can be fitted by power laws $F_{DFA} \sim n^{H}$. 
 Cross-correlations  are generated   
 since  we choose the error term to be 
 equal for both time series: $\eta_i = \eta'_i$, where 
 $\eta_i$ corresponds to $\{y_i\}$ and  $\eta'_i$ corresponds to $\{y'_i\}$. 
  When cross-correlations are present, 
the same weights are responsible for power-law  cross-correlations
between  $\{y_i\}$ and $\{y'_i\}$. For $n >> 1$ we find  
 $F_{DXA}(n) \approx n^{\lambda}$ [see Eq.~(4)], where $\lambda = 0.73$. 
 This example illustrates the relation:  
 $\lambda \approx  (H + H') / 2$.  
If we choose the error terms   $\eta'_i = - \eta_i$, then $F_{DXA}^2(n)$
becomes negative for every $n$. For that case the cross-correlation function
$X(n)$ becomes also negative. 
(b) Detrended covariance $F_{DXA}^2(n)$  of Eq.~(4). We generate  
two pairs of two  ARFIMA processes, where for each pair 
the time series are power-law auto-correlated, but not
cross-correlated, 
since each ARFIMA is generated by  its own error term.  
The  fluctuations, both positive and negative, 
 indicate  that 
	 two time series are not power-law cross-correlated with
	 an unique exponent, but 
	 either short-range cross-correlated or not
	  at all cross-correlated.    
} 
    \label{fig.2}
  \end{figure}

\newpage

\begin{figure}
    \narrowtext
    \centerline{
      \epsfysize=0.5\columnwidth{{\epsfbox{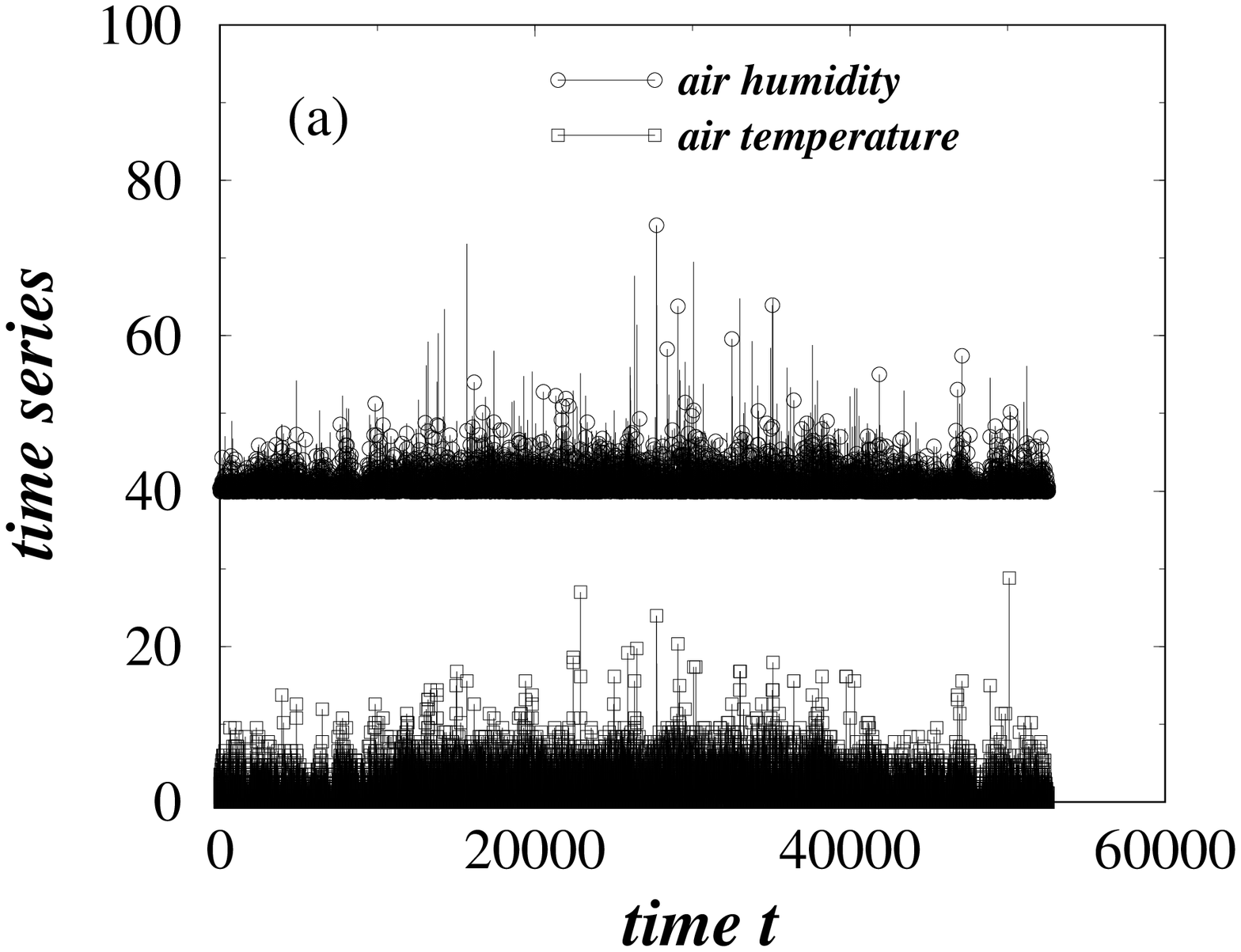}}} }
  \end{figure}
\begin{figure}
    \narrowtext
    \centerline{
      \epsfysize=0.5\columnwidth{{\epsfbox{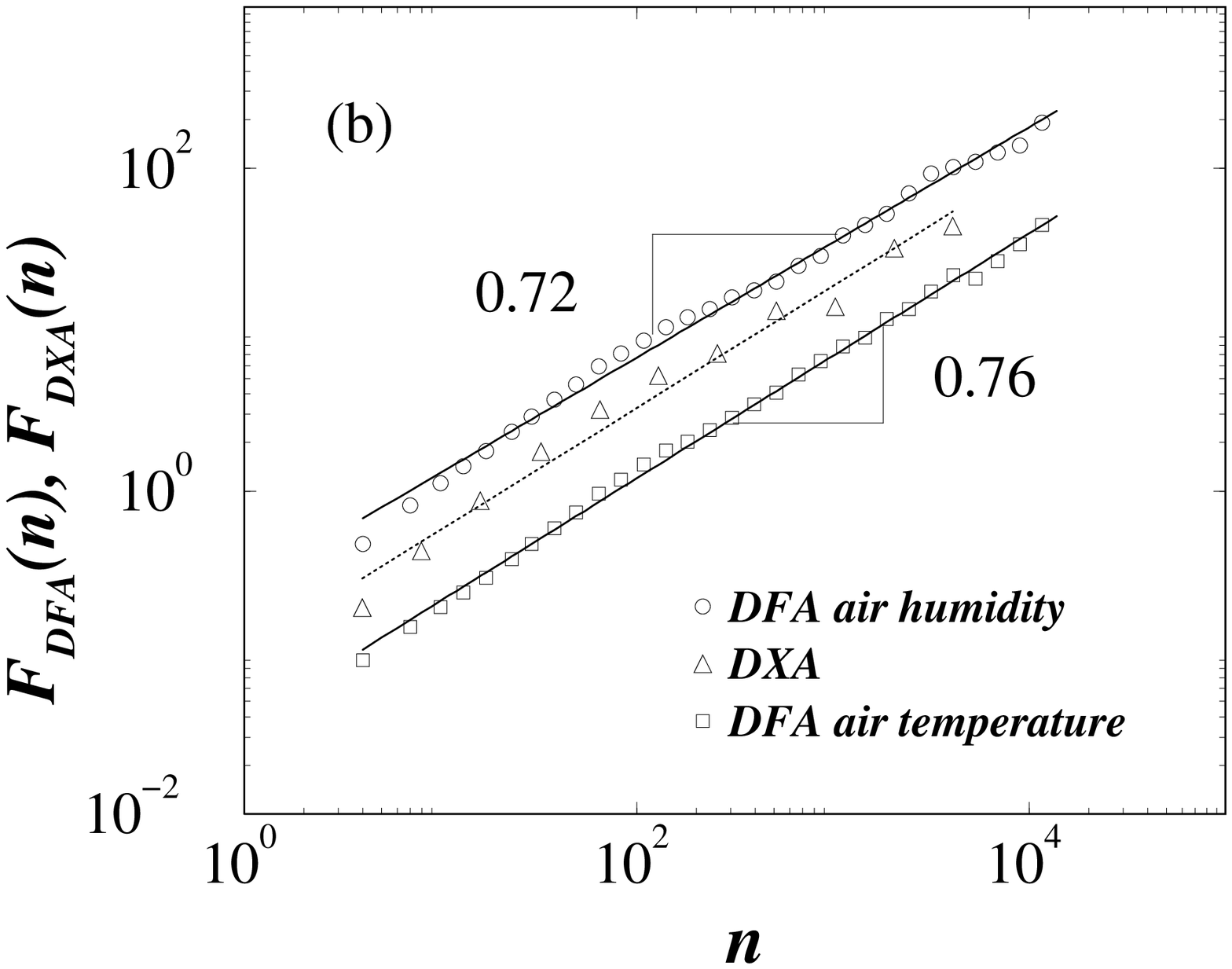}}}}
\caption{Power-law auto-correlations and cross-correlations 
in successive differences of air humidity $\{y_i\}$ 
and air temperature $\{y'_i\}$, recorded each 10 minutes. 
 (a) Time series 
of their absolute values, $\{|y_i|\}$ and $\{|y'_i|\}$.   
We find that both time series shows  sudden bursts of large changes. 
(b) The rms of detrended  variance $F_{DFA}(n)$  together with 
detrended  covariance $F_{DXA}(n)$.
 We find that DFA curves  of $\{|y_i|\}$ and
$\{|y'_i|\}$ and DXA curve  are very similar, 
and can be approximated with  power laws  $F_{DFA}(n) \sim n^{H}$ 
 with scaling exponents $H=0.72$ and $H'=0.76$, and 
$F_{DXA}(n) \sim n^{\lambda}$  with $\lambda=0.75$.  
}
    \label{fig.5}
  \end{figure}

\newpage

\begin{figure}
    \narrowtext
    \centerline{
      \epsfysize=0.5\columnwidth{{\epsfbox{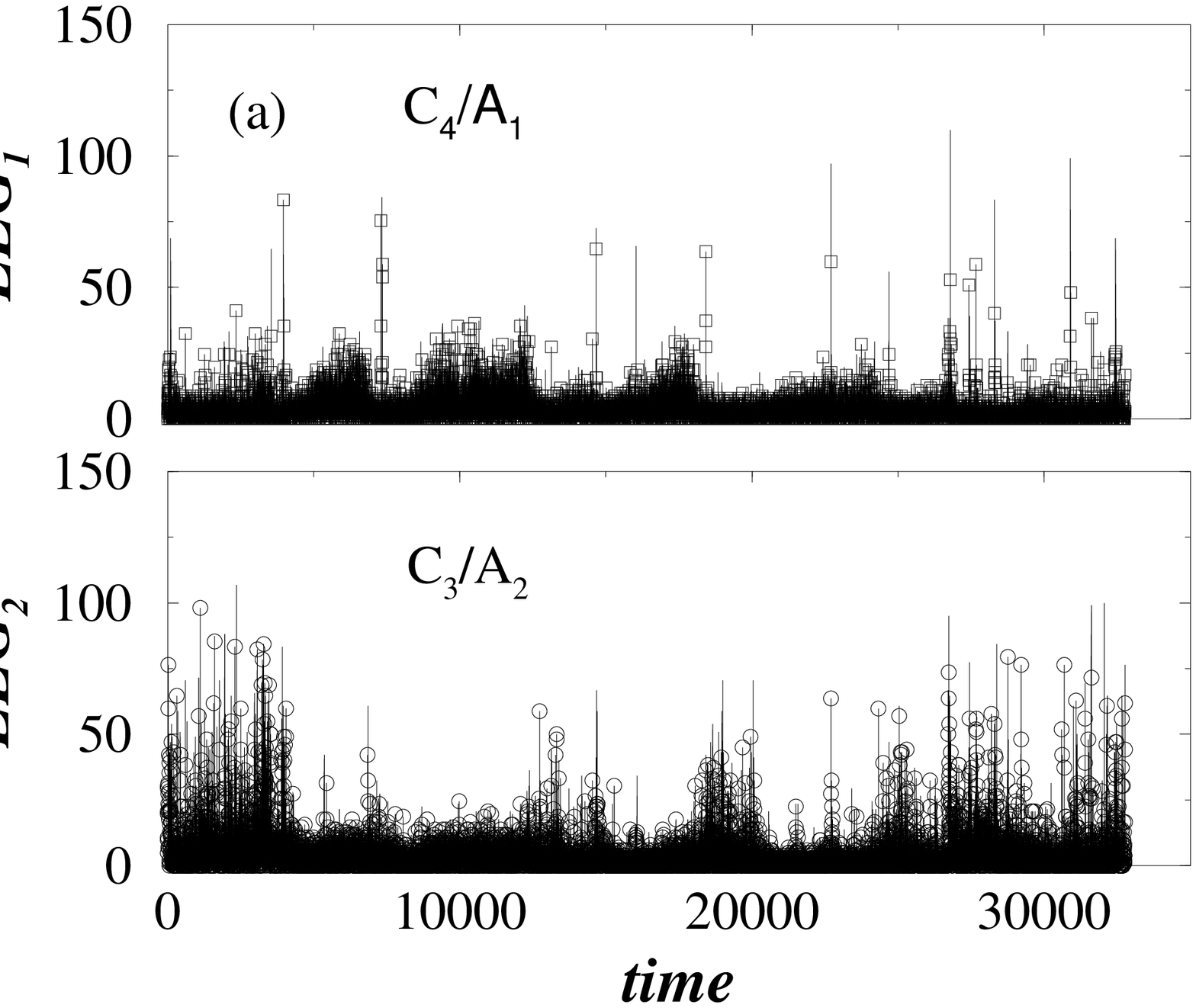}}} }
  \end{figure}
\begin{figure}
    \narrowtext
    \centerline{
      \epsfysize=0.5\columnwidth{{\epsfbox{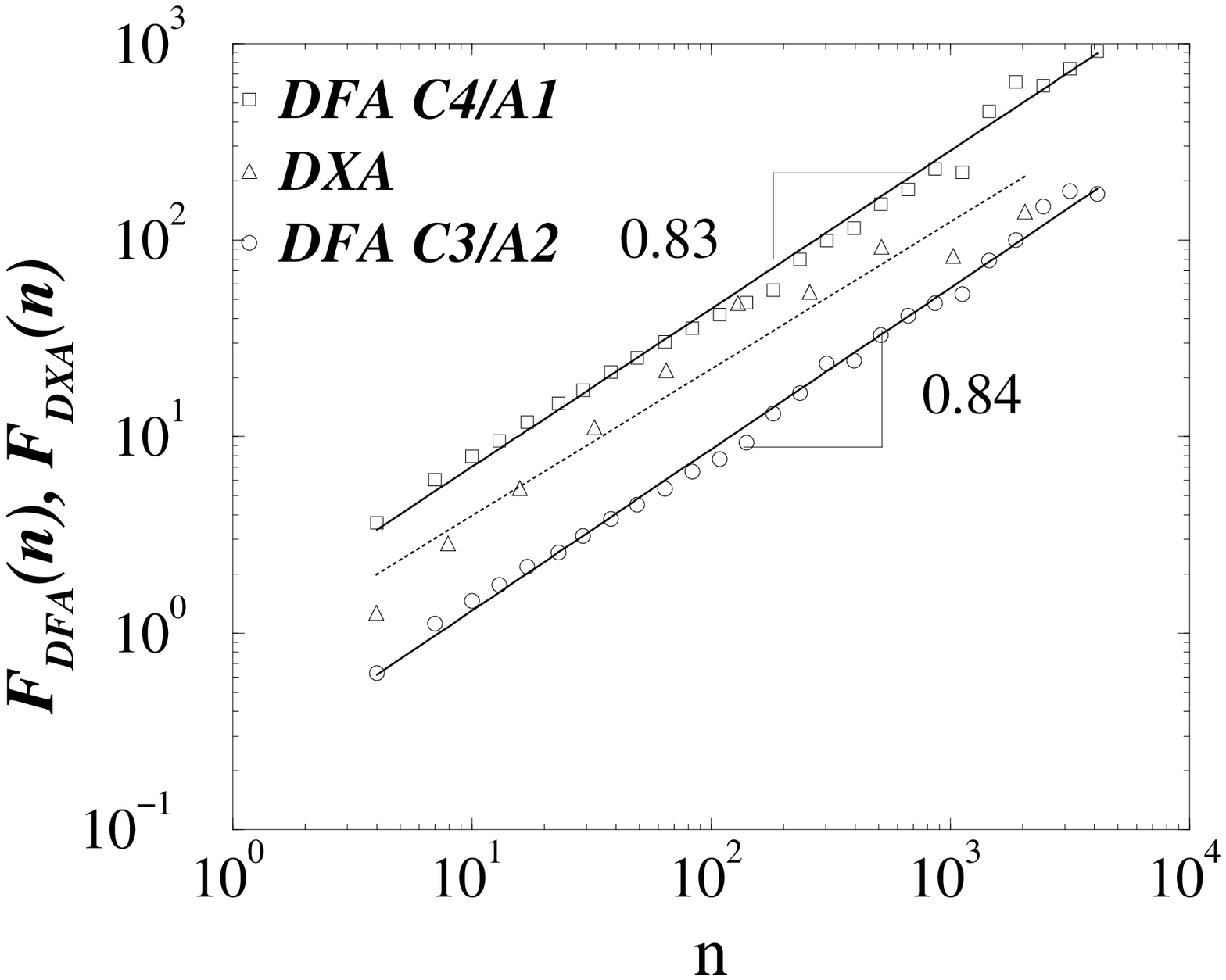}}}
    }
\caption{Power-law auto-correlations and cross-correlations  between two 
different EEG time series: $C4/A1$   $(\{y_i\})$ and $C3/A2$) 
$(\{y'_i\})$  with 32,000 data
  points, recorded every second. 
 (a)   Time series 
of their absolute values, $\{|y_i|\}$ and $\{|y'_i|\}$.   
(b) Rms of detrended  variance $F_{DFA}(n)$  together with 
detrended  covariance   $F_{DXA}(n)$. 
We find that $F_{DFA}(n)$  of $\{|y_i|\}$ and
$\{|y'_i|\}$ and DXA curve, $F_{DXA}(n)$  are very similar, 
and can be approximated with  power laws 
$n^{\alpha} (n^{\lambda})$ with scaling exponents $\alpha=0.83$ and $\alpha=0.84$,
$\lambda=0.84$, respectively. 
}
    \label{Physio}
  \end{figure}

\begin{figure}
    \narrowtext
    \centerline{
\epsfysize=0.7\columnwidth{{\epsfbox{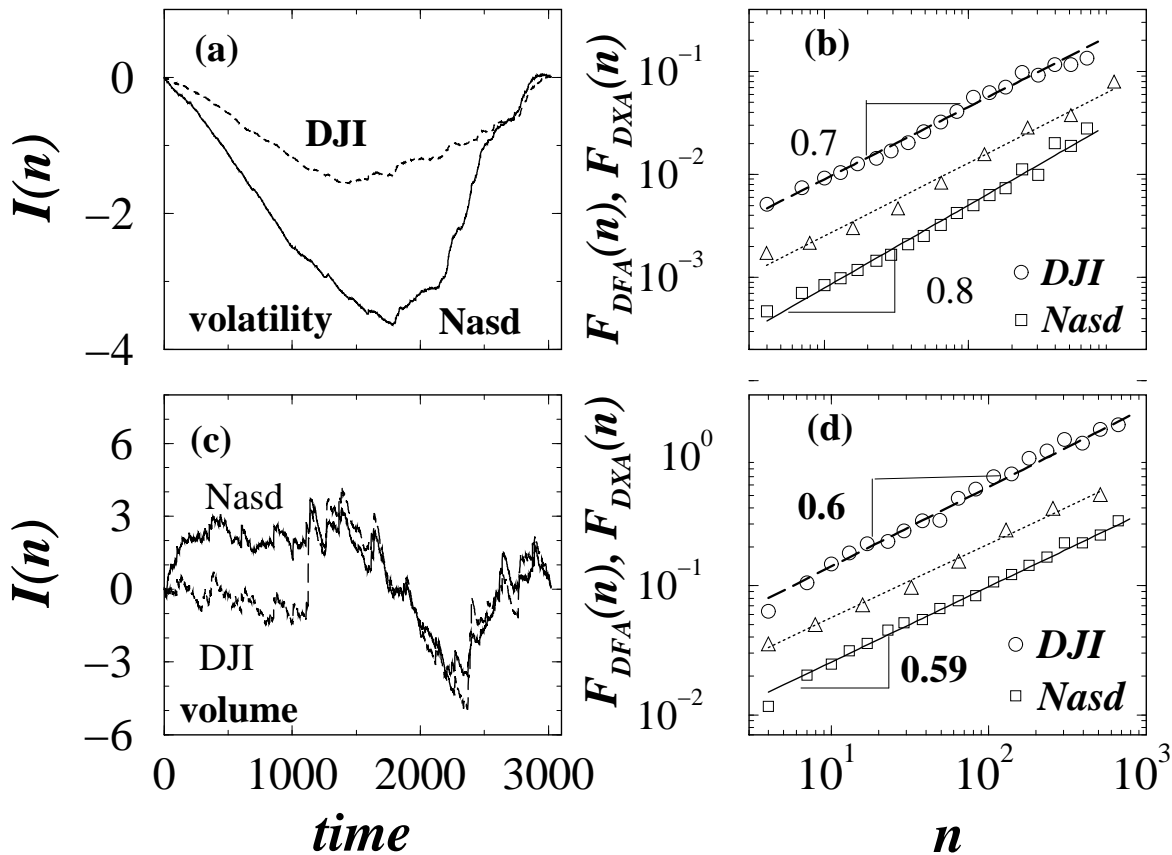}}}}
\vspace*{.9cm}
\caption{Long-range auto-correlations and 
cross-correlations in absolute values of price changes and
 trading volume  for  both Dow Jones Index and Nasdaq Index, 
 recorded daily, in the period from
July 1993 to November 2003. (a) Integrated profiles 
$I(n) \equiv \sum_{i=1}^n (|y_i| - \overline{|y_i|} )$ of the 
time series of absolute values
 $\{|y_i|\}$ and  $\{|y'_i|\}$ 
of logarithmic changes in price  for 
Dow Jones and  Nasdaq, and (c) the corresponding absolute  values 
 $\{|z_i|\}$ and $\{|z'_i|\}$   
for trading volume changes. 
 (b) For  price,   
 rms of the detrended variance $F_{DFA}(n)$  
 curves for each $\{|z'_i|\}$ and $\{|z'_i|\}$, and also  
the rms of the detrended covariance, $F_{DXA}(n)$. 
(d) Three curves represents the
same measures but for  ``volume'' (absolute of trading volume changes).  
 For all DFA and DXA curves 
we find  both power-law  auto-correlations and power-law cross-correlations. 
Power-law cross-correlations between Nasdaq and Dow Jones indices imply 
 that current price changes of Nasdaq depend on  its previous changes 
but also on previous price changes of Dow Jones. For trading volumes
 we also analyze time series $\{z_i\}$ and $\{z'_i\}$, and we find 
 that the DFA and DXA analyses show the anti-correlated behavior 
  with DFA exponents $\alpha=0.07$, 
 $\alpha=0.11$, and DXA exponent 
 $\lambda = 0.04$. 
}
\label{fig.7}
\end{figure} 

\begin{references}
\bibitem{power} The power spectrum  is the Fourier transform 
of the auto-correlation function. Similarly, for   
cross-correlations in the limit of small values of $f$, 
  $S(f) \sim f^{-\beta_{\times}}$  with 
  $X(n) \sim n^{-\gamma_{\times}}$  
 where $S(f) \approx  2 \sum_{n=1}^{\infty} 
 n^{-\gamma_{\times}} cos(2\pi fn)$ we obtain  
  $\beta = 1 - \gamma_{\times}$ 
by using the Taylor expansion  $(1-y)^{-1+\gamma_{\times}} = 1 +  
\sum_{n=1}^{\infty}
[ (1-\gamma_{\times})...(n-\gamma_{\times}) / \Gamma(n+1)  ] y^n$ and
a definition of $\Gamma$ function for $n>>1$, 
$\Gamma(z) \approx \Gamma(n+1) n^z / [ z(z+1)...(z+n)]$.
\bibitem{cutoff} We introduce a cutoff length $\ell = 10^4$ in
    our numerical simulations, and we let the sum run from 1 to $\ell$,
     i.e., we set $a_j=0$ for
    $j \ge \ell$.
\bibitem{artificial} Suppose one generates a sequence of N random numbers
 and then rerun the program  in order to obtain a new sequence. The same 
 sequence will be obtained and thus artificial cross-correlations will be
 created.
\bibitem{www} The data are publicly available at pdw.bic-gh.de.
\bibitem{remark} When $F_{DXA}^2(n)$ is negative for every $n$, one
may present -$F_{DXA}^2(n)$ vs. $n$ in log-log plot.
\end{references}
\end{document}